\documentclass[12pt,preprint]{aastex63}
\usepackage{amsmath}
\usepackage{verbatim}
\newcommand{\eqb}{\begin{equation}}
\newcommand{\eqe}{\end{equation}}

\begin{document}

\title{Fast radio bursts from reconnection in magnetar magnetosphere}
\author{Yuri Lyubarsky}
 \affil{Physics Department, Ben-Gurion University, P.O.B. 653, Beer-Sheva 84105, Israel}
\date{\today}
\begin{abstract}
The nearly 100\% linear polarization has been reported for a few fast radio bursts. This finding
places severe limits on the emission mechanism. I argue that the completely polarized radiation
could be generated in the course of relativistic magnetic reconnection in the outer magnetosphere
of the magnetar. At the onset of the magnetar flare, a large scale magnetic perturbation forms a
magnetic pulse, which propagates from the flare cite outwards. The pulse strongly compresses
magnetospheric plasma and pushes it away. The high-frequency MHD waves are generated when the
magnetic pulse reaches the current sheet separating, just beyond the light cylinder, the oppositely
directed magnetic fields. Coalescence of magnetic islands in the reconnecting current sheet
produces magnetosonic waves, which propagate away on the top of the magnetic pulse and escape in
the far zone of the wind as radio waves polarized along the rotational axis of the magnetar. I
estimate parameters of the outgoing radiation and show that they are compatible with the observed
properties of FRBs.

\end{abstract}

\section{Introduction}

Extraordinary properties of fast radio bursts (FRBs) impose high requirement on the emission
process. Two mechanisms are mostly discussed in the literature, the curvature emission of "bunches"
in a neutron star magnetosphere
\citep{Cordes_Wasserman16,Kumar17,Gisellini_Locatelli18,Katz18,Yang_Zhang18,Lu_Kumar19,Wang_etal19}
and the synchrotron maser emission
\citep{Lyubarsky14,Beloborodov17,Beloborodov19,Gisellini17,Waxman17,Gruzinov_Waxman19,Metzger19,Metzger_etal20}.

The available models of the curvature emission just assume that "bunches" with the required
properties exist in the magnetosphere; the emission of these bunches is estimated by applying
formulas of vacuum electrodynamics. Even leaving aside the question of how the bunches are formed
and whether they are able to survive in spite of an immense repulsive potential, one could not
ignore the fact that their radiation properties are strongly affected by the plasma
\citep{Gil_etal04}.
The required large charge of the bunches implies a very high plasma density, such that the plasma
frequency, $\omega_p$, is well above the emission frequency, $\omega$. Then, if the bunches move
together with the plasma, they are completely shielded and do not radiate at all. If they move with
respect to the plasma such that their velocity of is out of the range of the plasma thermal
velocities, their emission rate is only a fraction $(\omega/\omega_p)^2$ of the vacuum emission
rate, so that at the inferred parameters, the curvature emission is suppressed by many orders of
magnitude. The models ignoring plasma effects could not be considered as physically viable.

The synchrotron maser could provide, at reasonable conditions, the observed radiation power.
However, this mechanism is hardly compatible with the  100\% linear polarization observed in a few
FRBs \citep{Gijjar18,Michili18,oslowski19}. The synchrotron emission is totally polarized only in
the plane of the rotating particles; the out of the plane emission has an electric field component
parallel to the magnetic field so that the polarization of the total emission never reaches 100\%.
In the simulations of \citet{Gallant92}, a close to 100\% polarization  of the synchrotron mazer
emission from a relativistic shock has been observed however, this result seems to be an artefact
of 2D simulations. Namely, their simulation domain was resolved in the direction of the normal to
the shock front and in the direction of the magnetic field (parallel to the shock plane) so that
all parameters were homogeneous in the direction perpendicular to the magnetic field in the shock
plane. This means that radiation was generated in fact by currents homogeneous in this direction.
Such currents naturally produce 100\% linearly polarized emission. In effect, the synchrotron maser
emission is produced by charged bunches rotating in the magnetic field therefore their out of plane
emission should have a component polarized along the field. Only 3D simulations could provide a
realistic picture of the polarized synchrotron maser emission.


The totally linearly polarized radiation implies that one of the two polarizations modes is totally
suppressed in the source or on the way out. Such a condition is achieved if MHD waves with an
appropriate frequency are generated in a magnetosphere, where the magnetic energy density
 exceeds the plasma energy, including the rest energy. Any MHD perturbation
excites Alfven and fast magnetospnic (fms) waves. However, the group velocity of Alfven waves is
directed along the magnetic field so that only fms waves could propagate outwards, across the
magnetic field lines.
The condition that the energy in the magnetosphere is dominated by the magnetic energy is crucially
important because in this case, the fms waves are smoothly converted into vacuum electro-magnetic
waves when propagating towards the decreasing plasma density. 
In fms waves, the electric field is directed perpendicularly to the background
magnetic field, therefore they give rise to the radio emission polarized perpendicularly to the
background magnetic field.

High frequency MHD waves could be generated in the course of magnetic reconnection in a current
sheet separating oppositely directed magnetic fields. In the reconnection process, the current
sheet breaks into a system of linear currents, which are called magnetic islands because of their
islands-like appearance in 2D simulations. Inasmuch as the parallel currents are attracted one to
another, the islands continuously merge, see, e.g., the review by \citet{Kagan_etal15}. The merging
of two islands perturbs the surrounding magnetic field producing an fms pulse propagating radially
from the merging site.
This mechanism was recently proposed as a source of radio emission from the Crab and Crab-like
pulsars \citep{Uzdensky_Spitkovsky14,Lyubarsky19,Philippov19}.
In this paper, a model for FRBs is developed, in which the radio emission is produced by the
magnetic reconnection at the onset of the magnetar flare. A rotating magnetar continuously emits
relativistic, magnetized wind, similar to the pulsar wind. The oblique rotator introduces in the
wind zone an oscillating structure. 
In the equatorial zone of the wind, the field changes sign every half of period, therefore the
regions of opposite polarity are separated by corrugated current sheet. When a magnetar flare
shakes the magnetosphere, a strong magnetic perturbation produces a magnetic pulse that propagates
from the flare site outwards. Just beyond the light cylinder, the pulse catches up with the current
sheet, and the strong compression triggers violent reconnection, which gives rise to the radio
emission via merging of magnetic islands.

The paper is organized as follows. In sect.\ 2, I estimate parameters of the magnetic pulse  and
the characteristic frequency and the total energy of the MHD waves generated when the pulse
triggers reconnection in the equatorial current sheet. In sect.\ 3, parameters of the magnetar wind
are presented and propagation of the magnetic pulse in the wind zone is described. In sect.\ 4, I
consider escape of the MHD waves from the magnetic pulse and their conversion to the radio
emission. Conclusions are presented in sect.\ 4.

\section{Generation of fms waves}

At the onset of the magnetar flare, the twisted magnetic configuration  becomes unstable, which
yields a sudden rearrangement of the magnetic field (see, e.g., the recent review by
\citealt{Kaspi_Beloborodov17} and references therein).
The rapid motion of overtwisted magnetic field line tubes produces a large-scale MHD perturbation
that propagates outwards, opens the magnetosphere and propagates further out into the magnetar wind
\citep{Parfrey_etal13,Carrasco_etal19}. The formation and propagation of such a perturbation, a
magnetic pulse, has already been discussed by
\citet{Lyubarsky14,Beloborodov17,Beloborodov19,Metzger19,Metzger_etal20} in the context of an FRB
production by a magnetar flare. In this paper, I show that the FRB may be produced in the course of
magnetic reconnection triggered, just outside the light cylinder, by the magnetic pulse. Namely,
GHz magnetsonic waves are generated in the course of reconnection, and these waves give rise to
powerful radio emission\footnote{At the onset of the flare, the violent reconnection occurs also in
the inner magnetosphere however, all microscopic plasma parameters are very small there so that it
is not clear how the waves with the wavelength as large as 10 cm could be generated. In any case,
it is shown in Appendix A that such waves could not escape because of non-linear interactions.
Therefore in this paper, I consider the waves generated in the outer magnetosphere.}.

The outward propagating perturbation is in fact an fms pulse.
The amplitude of the pulse may be conveniently presented as
\eqb
B_{\rm pulse}=\sqrt{\frac{L_{\rm pulse}}c}\frac 1R=3.8\cdot 10^8\frac{L_{\rm
pulse,47}^{1/2}}{P}\frac{R_L}R\, G;
 \label{Bpulse}\eqe
where $L_{\rm pulse}$ is the isotropic luminosity associated with the pulse,  $R_L=4.8\cdot 10^9P$
cm the light cylinder radius, $P$ the magnetar period. Here and hereafter I employ the short-hand
notation $q_x=q/10^x$ in cgs units, e.g., $L_{47}=L/(10^{47} \rm erg\cdot s^{-1})$. The total
energy in the pulse is
\eqb
{\cal E}=L_{\rm pulse}\tau=10^{44}L_{\rm pulse,47}\tau_{-3}\,\rm erg,
 \label{energy-total}\eqe
 where $\tau=l/c$, $l$ is the width
of the pulse. The duration of the observed FRB will be shown to be about $\tau$.

Just beyond the light cylinder,  the pulse enters the magnetar wind.  The pulse propagates through
the wind practically with the speed of light.  The non-linear theory of the pulse propagation is
presented in Appendix B; here I give only a short outline. The plasma is squeezed in the pulse and
pushed forward; the plasma velocity with respect to the wind is estimated as the velocity of the
zero electric field frame, $v'=E'/B'=B'_{\rm pulse}/(B'_{\rm wind}+B'_{\rm pulse})$, where $B_{\rm
wind}$ is the magnetic field of the wind; the prime is referred to quantities in the wind frame.
The corresponding Lorentz factor is
\eqb
\Gamma=\sqrt{\frac{B'_{\rm pulse}}{2B'_{\rm wind}}}=\frac 12\sqrt{\frac{B_{\rm pulse}}{B_{\rm
wind}}}=100\frac{L_{{\rm pulse,}47}^{1/4}P}{\mu_{33}^{1/2}}.
 \label{Gamma}\eqe
Here the magnetic field of the wind is expressed via  the magnetic momentum of the magnetar, $\mu$:
\eqb
B_{\rm wind}=B_L\frac{R_L}{R};\quad B_L=\frac{\mu}{R_L^3}=9\cdot 10^3 \frac{\mu_{33}}{P^{3}}\,\rm
G.
 \label{Bwind}\eqe
 Note that $\Gamma$ is the Lorentz factor of the plasma within the pulse. The pulse itself, i.e.
the waveform, moves with so large Lorentz factor that one can safely assume that the pulse moves
with the speed of light; see Appendix B.

In the magnetar wind, a current sheet separates two magnetic hemispheres, the shape of the sheet
has being likened to a ballerina's skirt. When the magnetic pulse arrives at the sheet, the sharp
acceleration and compression cause the violent reconnection. 
Moreover, the current sheet is destroyed by the Kruskal-Schwarzschild instability
\citep{Lyubarsky10,Gill_etal18}, which is the magnetic counterpart of the Rayleigh-Taylor
instability, so that the field line tubes with the oppositely directed fields fall into the
magnetic pulse forming multiple small current sheets 
scattered over the body of the pulse. Within each of the small current sheets, the reconnection
process occurs via formation and merging of magnetic islands, which gives rise to an fms noise.
This noise is converted, as it will be shown below, into the radio emission. Since the sources of
the noise are distributed in the body of the magnetic pulse but not concentrated at the front part,
the duration of the observed radiation burst is of the order of the duration of the pulse,
$\tau\sim l/c$.  Let us estimate the total energy of the generated waves and their characteristic
frequencies.

{\it The emitted energy.} The total reconnecting magnetic flux may be estimated as $B_LR_L^2$.
Within the pulse, the stripe with the oppositely directed field is compressed $B_{\rm pulse}/B_{\rm
wind}$ times.
Then the total energy of annihilated fields is roughly
\eqb
\varepsilon\sim\left(B_{\rm pulse}/B_{\rm wind}\right)B^2_LR_L^3.
\eqe
According to the simulations by \citet{Philippov19}, the fraction $f\sim 0.01$ of the reconnecting
magnetic energy is emitted in the form of fms waves. Then the energy of the radio burst may be
estimated as
\eqb
\varepsilon_{\rm FRB}=f{\varepsilon}=3.8\cdot 10^{39}\frac{f_{-2}\mu_{33}L_{{\rm
pulse,}47}^{1/2}}{P}\,\rm erg.
 \label{energy}\eqe

{\it The characteristic frequencies.} The characteristic frequency of the emitted waves is
determined by the collision time of two merging islands. The size of the islands is 10-100 times
larger than the width of the current sheet, $a'$, \citep{Philippov19} therefore one can write
 \eqb
\omega'=\frac c{\xi a'},
\eqe
where $\xi\sim 10-100$. Here and hereafter, quantities measured in the plasma comoving frame are
marked by prime.
The width of the sheet may be estimated as it has been done for the pulsar current sheet
\citep{Lyubarskii96,Uzdensky_Spitkovsky14}. Within the sheet, the pressure of the external magnetic
field is balanced by the pressure of "hot" pairs,
\eqb
N'm_ec^2\gamma'_T=\frac{B'^2_{\rm pulse}}{8\pi}.
\eqe
Here $N'$ is the pair density within the sheet, $\gamma'_T$ the characteristic Lorentz factor of
the pairs, $m_e$ the electron mass.  The reconnection occurs via the collisionless tearing
instability therefore the width of the sheet, $a'$, is of the order of a few  particle Larmor
radii,
\eqb
a'=\zeta\frac{m_ec^2\gamma'_T}{eB'_{\rm pulse}},
\eqe
where $\zeta$ is  a few. The synchrotron cooling time is very short at the inferred parameters
therefore within the sheet, the reconnection energy release is balanced by the synchrotron cooling,
\eqb
\epsilon\frac{B'^2_{\rm pulse}}{4\pi}c=N'\sigma_T\frac{B'^2_{\rm pulse}}{4\pi}c\gamma'^2_Ta'.
\eqe
Here $\epsilon\sim 0.1$ is the reconnection rate. Eliminating from the last three equations
$\gamma_T$ and $N$ in favor of $a$, one gets
\eqb
a'=\left(\frac{3\epsilon\zeta}{r_e}\right)^{1/2}\left(\frac c{\omega'_B}
\right)^{3/2},
\eqe
where $r_e$ is the classical electron radius, $\omega'_B=eB_{\rm pulse}/m_ec$ the cyclotron frequency. 
Now the emitted frequency in the observer's frame may be estimated as
\eqb
\nu=2\Gamma\frac{\omega'}{2\pi}=\frac 1{\pi\xi} \left(\frac{r_e}{3\epsilon\zeta
c\Gamma}\right)^{1/2}\omega_B^{3/2}
=3\frac{\mu_{33}^{1/4}L_{\rm pulse,47}^{5/8}}{\xi_1\zeta_1^{1/2}\epsilon_{-1}^{1/2}P^2}\,\rm
GHz.
 \label{frequency}\eqe

The emitted fms waves propagate on the top of the magnetic pulse and escape from it far away from
the magnetar. On the way out, different processes could affect their propagation. Let us consider
the fate of the fms waves and find at what conditions they eventually escape as radio waves.

\section{The magnetar wind and propagation of the magnetic pulse}

In order to study propagation and escape of the fms waves, let us consider the properties of the
pulsar wind and of the magnetic pulse, within which the waves are generated.
The particle flux in the wind from a strongly twisted magnetar magnetosphere could reach
\citep{Beloborodov19}
\eqb
\dot{N}\sim{\cal M}\frac{c\mu}{eR_LR_{\pm}},
\eqe
where ${\cal M}\sim 10^3$ is the pair multiplicity, $R_{\pm}\sim 5\cdot 10^6\mu^{1/3}_{33}$ cm the
distance from the star  where the magnetic field falls to $10^{13}$ G such that the pair production
stops. With the above particle flux, the magnetization parameter of the magnetar wind, which is
defined as the ratio of the Poynting flux to the rest mass energy flux in the wind, is
 \eqb
\eta=\frac{B_L^2R_L^2}{\dot{N}m_ec}=2.5\cdot 10^4\frac{\mu_{33}^{4/3}}{{\cal M}_3P^3}.
 \eqe
The magnetization parameter is in fact the maximal Lorentz factor achievable by the wind if the
magnetic energy is converted to the kinetic energy.

Beyond the light cylinder, the strongly magnetized wind is accelerated
 linearly with the distance, $\gamma\sim R/R_L$, until it reaches the fast
magnetosonic point, $\gamma\sim\eta^{1/3}$. Beyond the fast magnetosonic point, the wind
accelerates very slowly, $\propto(\ln R)^{1/3}$ \citep{Beskin98}, if there is no dissipation. In
this case, one can take the Lorentz factor of the wind in the far zone, $R\gg \eta^{1/3}R_L$, to be
 \eqb
\tilde{\gamma}_{\rm wind}=3\eta^{1/3}=90\frac{\mu_{33}^{4/9}}{{\cal M}^{1/3}_3P}.
 \eqe
The magnetic dissipation could lead to a gradual acceleration of the wind in the equatorial belt,
where the magnetic field changes sign every half of period
\citep{Lyubarsky_Kirk01,Kirk_Skjeraasen03}; then the Lorentz factor of the wind exceeds
$\tilde{\gamma}_{\rm
wind}$ and may even reach $\eta$. 

 The wind is terminated at a strong shock formed when the dynamic pressure of the wind is balanced by the
plasma pressure within the nebula surrounding the magnetar. The radius of the wind termination
shock is estimated as:
 \eqb
R_s=\left(\frac{4\pi^3\mu^2}{P^4c^3p}\right)^{1/2}=1.2\cdot
10^{15}\frac{\mu_{33}}{P^2p_{-4}^{1/2}}\, \rm cm.
 \label{term-shock}\eqe
Here I normalized the pressure within the nebula, $p$, to $10^{-4}$ dyn/cm$^2$; such a pressure is
found in the nebula surrounding the repeater FRB 121102 \citep{Beloborodov17}.

 The magnetic pulse moves through the magnetar wind as a
propagating wave (see Appendix B) so that the plasma enters the pulse through the front part and
eventually leaves it through the rear part.   Within the pulse, the plasma moves with respect to
the wind with the Lorentz factor (\ref{Gamma}).  In the lab frame, the Lorentz factor of the plasma
within the pulse is
  \eqb
\gamma_{\rm pulse}=2\gamma_{\rm wind}\Gamma=\gamma_{\rm wind}\sqrt{\frac{B_{\rm pulse}}{B_{\rm
wind}}}= 1.8\cdot 10^4\frac{L^{1/4}_{\rm pulse, 47}}{\mu_{33}^{1/18}{\cal
M}^{1/3}_3}\frac{\gamma_{\rm wind}}{\tilde{\gamma}_{\rm wind}}.
  \label{gamma_pulse}\eqe
Inasmuch as the plasma acquires a very high Lorentz factor, it is dragged within the pulse to a
large distance and is substituted by the wind plasma very slowly.

The pulse picks up the magnetospheric plasma at the region where the magnetic flux in the pulse,
 \eqb
\Phi_0= B_{\rm pulse}lR=l\sqrt{\frac{L_{\rm pulse}}c},
 \label{Phi0}\eqe
  becomes equal to the magnetospheric
flux, $BR^2$, where $B=\mu/R^3$. This occurs at the distance
 \eqb
R_0=\frac{\mu}{\tau\sqrt{L_{\rm pulse}c}}=2\cdot 10^7\frac{\mu_{33}}{\tau_{-3}L^{1/2}_{\rm
pulse,47}}\,\rm cm
 \label{R0}\eqe
 i.e, well within the magnetosphere.
Let us estimate the amount of plasma available in this region.

  The magnetar magnetosphere is
filled by an electron-positron plasma generated in cascades induced by slow untwisting of
magnetospheric field lines \citep{ThompsonLyutikovKulkarni02,
BeloborodovThompson07,Thompson08,Beloborodov13a}. The pair injection rate is determined by currents
in the magnetosphere. In a strongly twisted magnetosphere, the magnetic field lines with the apex
radii $\sim R$ carry the current $j\sim cB/R$ because the "toroidal" component of the magnetic
field could not exceed, by stability considerations, the poloidal one. The theory of pair
production in magnetars predicts the plasma density
 \eqb
N\sim{\cal M} j/ec\sim{\cal M} B/eR,
 \label{density}\eqe
where ${\cal M}\sim 100-1000$ is the pair multiplicity \citep{Beloborodov13a}. Then the total
number of particles in the region $R\sim R_0$ is estimated as
 \eqb
{\cal N}\sim NR^3=\frac{{\cal M}\mu}{eR_0}.
 \eqe
The pulse picks up these particles and drags them out of the magnetosphere. One can estimate the
ratio of the Poynting to the kinetic energy fluxes in the pulse in the far zone as
 \eqb
\sigma_{\rm pulse}=\frac{\cal E}{m_ec^2{\cal N}\gamma_{\rm pulse}}=4.5\cdot
10^{10}\frac{L^{1/4}_{\rm pulse, 47}\mu^{1/18}_{33}}{{\cal M}_3^{2/3}} \frac{\tilde{\gamma}_{\rm
wind}}{\gamma_{\rm wind}}.
 \label{sigma-pulse}\eqe
 This estimate was obtained assuming that the plasma density in the magnetosphere remained the
 same as in a quiet state. 
However, one can expect that already at the onset of the flare, some amount of plasma is produced.
Then $\sigma_{\rm pulse}$ decreases but still remains extremely high.

The new plasma enters the pulse through the front part together with the new magnetic flux. Since
the magnetar wind runs away with a relativistic velocity, the newly accumulated flux is estimated
as
 \eqb
\Phi=\int B_{\rm wind}R\left(c-v_{\rm wind}\right)dt=\int \frac{B_{\rm wind}R}{2\gamma^2_{\rm
wind}}dR.
 \eqe
In the striped part of the wind (in the equatorial belt), only the average magnetic flux is
accumulated, the alternating part of the field being annihilated either in the wind before the
pulse arrives (and then the wind is accelerated) or when compressed within the pulse. Therefore one
has to substitute $B_{\rm wind}$ in this estimate by $\xi B_{\rm wind}$, where $B_{\rm wind}$ is
given by eq.\ (\ref{Bwind}) and the latitude dependent coefficient $\xi<1$ is the ratio of the mean
to the total field.
Now the accumulated flux may be presented as
 \eqb
\Phi=\frac{\xi B_LR_LR}{2\gamma^2_{\rm wind}}.
 \eqe
 The new flux and plasma form a layer in the front part of the pulse. The width of the layer, $\Delta$,
 is presented as
 \eqb
 \frac{\Delta}l=\frac{\Phi}{\Phi_0}=0.2\frac{\xi{\cal
 M}_3^{2/3}\mu_{33}^{1/9}}{\tau_{-3}L_{\rm pulse,47}^{1/2}}
 \left(\frac{\tilde{\gamma}_{\rm wind}}{\gamma_{\rm wind}}\right)^2R_{15}.
 \label{Delta}\eqe
One can expect that at the distances of interest, at least a part of the alternating field is
annihilated in the wind so that $\gamma_{\rm wind}>\tilde{\gamma_{\rm wind}}$. Taking this into
account, and also that $\xi<1$ (at the equator, $\xi=0$), one can expect that when the pulse
arrives at the wind termination shock at the radius (\ref{term-shock}), the layer with the new
plasma and flux fills only a small part of the pulse, the mean body of the pulse being still filled
with the plasma picked up in the magnetosphere.

The pulse may be considered as a propagating wave until the shock is formed due to the non-linear
steepening of the pulse. The non-linear steepening occurs because the wave velocity depends on the
local density and magnetic field, which vary across the pulse. In a highly magnetized plasma, the
non-linearity is weak and therefore the non-linear steepening scale
 in the comoving plasma frame is large (\citealt{Lyubarsky03}, see also Appendix B)
 \eqb
  s'_{\rm steep}\sim \sigma_{\rm pulse} l'.
 \label{steepening}\eqe
Transforming to the lab frame and making use of eq. (\ref{gamma_pulse}), one finds that the shock
formation distance is
 \eqb
R_{\rm steep}=4\sigma_{\rm pulse}\gamma^2_{\rm pulse}l=4\cdot 10^{16}\sigma_{\rm pulse}\frac{L_{\rm
pulse,47}^{1/2}\tau_{-3}}{\mu_{33}^{1/9}{\cal M}^{2/3}_3}\left(\frac{\gamma_{\rm
pulse}}{\tilde{\gamma_{\rm pulse}}}\right)^2\,\rm cm.
 \label{steepening1}\eqe
 Taking into account that $\sigma_{\rm pulse}\gg 1$ (see above), one concludes that the shock
could hardly be formed while the pulse propagates within the magnetar wind.

 The above considerations assume implicitly that the mean field in the wind is roughly in the same
direction as the initial field of the pulse. Taking into account that the mean field has opposite
signs in two hemispheres, one can imagine a situation when the accumulated flux is opposite to the
flux in the pulse.  In an ideal case, this does not affect the obtained results because one can
easily imagine a pulse within which the magnetic field changes sign such that a current sheet
separates domains of the opposite polarity. The electro-magnetic stress is a quadratic function of
the fields therefore a solution of ideal MHD equations is not affected if one reverses fields in
any bundle of the magnetic field lines and inserts an appropriate current sheet provided the plasma
in the sheet is light enough (so that the overall inertia is not affected). In reality, the current
sheet is unstable and one can expect that narrow magnetic tubes of the opposite polarity fall into
the main body of the pulse and dissipate there, as it was discussed in sect.\ 2. In this case, one
can expect that the overall magnetization of the pulse decreases and the pulse is accelerated
because the heated plasma expands transforming heat to the kinetic energy. The detailed analysis of
this process is beyond the scope of this paper. As it was discussed above, the pulse accumulates a
relatively small new flux before arrives at the wind termination shock or at the previously ejected
baryonic cloud. Therefore one can expect that the above results remain valid independently of the
relative magnetic polarity in the pulse and in the wind.

\section{Escape of the waves}

Fast magnetosonic waves are in fact sound waves in which longitudinal plasma oscillations are
maintained by the magnetic pressure. In the relativistic case, their phase velocity is found as
(e.g., \citealp{Appl_Camenzind88})
\eqb
\frac{\omega}k=c\sqrt{\frac{\sigma}{1+\sigma}},
 \label{fms-disp}\eqe
where $\sigma=B^2/4\pi\rho c^2$. The electric field in the wave, $\mathbf{E}=-\frac
1c\mathbf{v\times B}$, is directed perpendicularly to the propagation direction therefore this is a
transverse electro-magnetic wave. At $\sigma\to\infty$, the phase velocity of the wave goes to the
speed of light. The conductivity current in the wave is found by substituting a harmonic wave with
the above dispersion law into the Maxwell equations,
\eqb
\nabla\times\mathbf{E}=-\frac 1c\frac{\partial\mathbf{B}}{\partial t};\qquad \nabla\times
B=\frac{4\pi}c\mathbf{j}+\frac 1c\frac{\partial\mathbf{E}}{\partial t},
\eqe
and taking into account that in this wave, ${\mathbf{k\cdot E}}=0$. Then one finds that the ratio
of the conductivity to the displacement current is
\eqb
\frac j{i\omega E}=\frac 1{4\pi(1+\sigma)}.
 \label{displacement}\eqe
One sees that at $\sigma\to\infty$, the conductivity current vanishes so that when the plasma
density goes to zero, the fms wave becomes a vacuum electro-magnetic wave.

This could be easily understood as follows. Consider a vacuum electro-magnetic wave superimposed on
a homogeneous magnetic field such that the electric field of the wave is perpendicular to the
background field. A charged particle in such a system experiences drift motion in the crossed
electric and magnetic fields, the direction of motion being independent of charge. Therefore if a
small amount of plasma is added to the system, no electric current appears in the system, and the
wave propagates as in vacuum.

The fms waves propagate outwards on the top of the magnetic pulse, that runs through the magnetar
wind.  The pulse propagates practically with the speed of light. Recall that the velocity of the
pulse is the velocity of the waveshape; the plasma moves a bit slower, even though highly
relativistically (see eq.\ \ref{gamma_pulse}). The waveshape propagates with the fms velocity with
respect to the plasma so that the Lorentz factor of the waveshape is extremely large,
$\sim\gamma_{\rm pulse}\sqrt{\sigma_{\rm pulse}}$ (see appendix B). Therefore any deviations of the
pulse velocity from the speed of light are negligible. The pulse  is decelerated only when it
collides with a stationary or slowly moving plasma. This happens either at the wind termination
shock when the pulse enters the magnetar nebula \citep{Lyubarsky14}, or, in the case of repeaters,
when the pulse collides with the previously ejected baryonic material \citep{Beloborodov17,
Metzger19, Beloborodov19,Metzger_etal20}. When the pulse is decelerated, the fms waves escape as
radio waves. The observed duration of the FRB is determined by the width of the pulse and therefore
it is of the order of $\tau$.
 Different processes could affect these waves on the way out; let us consider them.

First of all, let us consider  {\it the cyclotron absorption}. Making use of eq. (\ref{Bpulse}),
one finds that the cyclotron resonance, $\omega'=\omega'_B\equiv eB'_{\rm pulse}/m_ec$, is reached
at the radius
 \eqb
 R_{\rm cycl}=5\cdot 10^{15} \frac{L_{\rm pulse, 47}^{1/2}}{\nu_9}\,\rm cm,
 \label{cycl_radius}\eqe
 which is comparable with the radius of the termination shock (\ref{term-shock}) and therefore in some cases, the cyclotron
absorption should be taken into account. The resonant cross section is
\eqb
\sigma_{\rm res}=4\pi^2 \frac{e^2}{m_ec}\delta(\omega'-\omega'_B).
\eqe
Now the cyclotron absorption depth is found as
 \eqb
\tau_{\rm cycl}=\int\sigma_{\rm res} N'dR'=
\frac{4\pi^2eN'}{\frac{dB'}{dR'}},
 \eqe
where $N'$ is the lepton density in the comoving frame. Making use of the estimate $dB'/dR'\sim
B'\gamma_{\rm pulse}/R$, one finds
 \eqb
\tau_{\rm cycl}\sim\frac{\pi\omega'_B R_{\rm cycl}}{\sigma_{\rm pulse}\gamma_{\rm
pulse}c}=\frac{\pi\omega R_{\rm cycl}}{2\sigma_{\rm pulse}\gamma_{\rm pulse}^2c}=1.6\cdot
10^7\,\frac{{\cal M}_3^{2/3}\mu^{1/9}}{\sigma_{\rm pulse}}\left(\frac{\gamma_{\rm
pulse}}{\tilde{\gamma_{\rm pulse}}}\right)^2,
 \label{cycl_depth}\eqe
where I used  eqs. (\ref{Bpulse}) and (\ref{gamma_pulse}) and the general relation for the ratio of
the Poynting to the kinetic energy flux $\sigma=\omega'^2_B/\omega'^2_p$.

One sees that the synchrotron absorption is weak if $\sigma_{\rm pulse}> 10^7$. According to the
estimate (\ref{sigma-pulse}), the initial magnetization of the pulse satisfies this constraint.
However, when the wind matter enters the pulse, the magnetization becomes significantly lower and
therefore the cyclotron absorption may be significant in the layer where the wind plasma is
accumulated. As it was argued above,
only a fraction of the pulse is filled by the wind matter at the cyclotron resonance
point (\ref{cycl_radius}), therefore only a fraction of the radiation is absorbed, the rest
escaping freely. Another option is that the waves escape the pulse before the cyclotron resonance
radius is reached. The last occurs when the radius of the magnetar wind terminations shock
(\ref{term-shock}) is smaller than the cyclotron resonance radius. The pulse is sharply decelerated
just beyond the termination shock \citep{Lyubarsky14}, and the waves escape into the nebula where
the magnetic field is significantly smaller than within the pulse so that $\omega>\omega_B$. In
repeating FRBs, the termination shock is pushed outwards by the previous bursts and therefore it
may be well beyond the cyclotron resonance radius. In this case, however, the magnetic pulse is
sharply decelerated when it enters the mildly relativistic baryonic material ejected by the
previous flare \citep{Metzger19,Beloborodov19,Metzger_etal20}. This occurs at relatively small
distances from the magnetar, $\sim 10^{12}-10^{14}$ cm, therefore the waves leave the pulse well
before the cyclotron resonance radius is reached.

Now let us consider {\it the non-linear interactions of MHD waves}, which could produce a wave
cascade that transfers the wave energy out of the observed spectral range.  The non-linear
interactions of force-free MHD waves were studied by \citet{Thompson_Blaes98} and
\citet{Lyubarsky19}. The strongest are the three-wave interactions: 
 \eqb
\omega=\omega_1+\omega_2;\quad \mathbf{k}=\mathbf{k}_1+\mathbf{k}_2.
 \eqe
In the force-free MHD, the dispersion relation for the fms waves is $\omega=ck$, and for the Alfven
waves $\omega=ck\vert\cos\theta\vert$, where $\theta$ is the angle between the background magnetic
field and the direction of the wave. Substituting these dispersion laws into the conservation laws
one sees that only interactions involving both types of waves are possible:
 \eqb
 {\rm fms}\rightleftarrows {\rm fms}+{\rm Alfven}\quad {\rm and}\quad
{\rm fms}\rightleftarrows {\rm Alfven}+{\rm Alfven}.
 \label{fms_decay}\eqe
 The rate of the interaction may be roughly estimated, in the zero electric field frame, as
\eqb
q'\sim\left(\frac{\delta B'}{B'}\right)^2\omega',
 \label{q}\eqe
where $\delta B$ is the amplitude of the waves, $B$ the background field.

 The induced decay of fms waves (\ref{fms_decay}) is possible only in the presence of Alfven waves.
However, they do not propagate across the magnetic field lines and therefore remain close to the
reconnection cites. Far from the reconnection cites, they  grow exponentially from small
fluctuations with the rate (\ref{q}), which requires many e-folding times. Therefore the condition
that the fms waves do not decay significantly may be written as
 \eqb
\tau_{\rm NL}\equiv   \int_{R_L}^R q'dt'  <10.
 \label{NLcondition}\eqe
  As soon as the pulse propagates  in the accelerating wind, the Lorentz factor of the plasma in the pulse
also increases linearly with the distance. The ratio $\delta B'/B'$ is independent of the distance
and the Lorentz factor of the flow however, the proper wave frequency decreases as $\gamma^{-1}\sim
r^{-1}$ and the proper time goes as $dt'\propto dr/r$. Therefore as long as the flow is
accelerated, $\tau_{\rm NL}$ is determined by the initial region $R\sim R_L$, where the wind is
still mildly relativistic and the pulse Lorentz factor is given by eq. (\ref{Gamma}). Taking into
account that the ratio $(\delta B'/B')^2$  is just the ratio of the fms energy (\ref{energy}) to
the total energy of the magnetic pulse (\ref{energy-total}), one gets
 \eqb
\tau_{\rm NL}\sim\frac{{\varepsilon}_{\rm FRB}}{\cal E}\frac{\omega R_L}{2\Gamma^2
c}=1.9\frac{f_{-2}\mu_{33}^2\nu_9}{L_{\rm pulse,47}P^2\tau_{-3}},
 \label{NL_depth}\eqe
which means that the fms signal is not affected significantly by the three-wave decay process in
the vicinity of the magnetar.

In the far zone, where the wind stops accelerating or accelerates slowly, the non-linear optical
depth begins to increase:
 \eqb
\tau_{\rm NL}=\frac{{\varepsilon}_{\rm FRB}}{\cal E}\frac{\omega R}{2\gamma^2_{\rm pulse} c}=
50\frac{f_{-2}\mu^{10/9}_{33}\nu_9{\cal M}^{2/3}_3}{L_{\rm
pulse,47}P\tau_{-3}P}\left(\frac{\gamma_{\rm wind}}{\tilde{\gamma_{\rm wind}}}\right)^2R_{15}.
 \eqe
One sees that at large enough distances from the magnetar, the non-linear wave-wave interaction
could have remove the wave energy from the observed spectral range. However, the MHD approximation
is already violated at these distances. Namely, the MHD approximation assumes that the proper wave
frequency is less than both the plasma frequency, $\omega'_p=(4\pi e^2 N'/m_e)^{1/2}$, and the
cyclotron frequency, $\omega'_B=eB'/mc$. Making use of the relation $\sigma=(\omega_B/\omega_p)^2$,
the ratio of the proper wave frequency to the plasma frequency in the pulse is estimated, with the
aid of eq. (\ref{Bpulse}), as
 \eqb
\frac{\omega'}{\omega'_p}=\frac{\omega'\sigma_{\rm pulse}^{1/2}}{\omega'_B}=0.2\sigma^{1/2}_{\rm
pulse}\frac{R_{15}\nu_9}{L_{\rm pulse, 47}^{1/2}}.
 \label{MHDcond}\eqe
One sees that the MHD approximation is violated (in fact the Alfven waves do not exist at
$\omega'>\omega'_p$) before the MHD non-linear interactions become significant if $\sigma_{\rm
pulse}$ exceeds a few dozens, which  is not a severe constraint.

One more important effect is the {\it non-linear steepening}, which could lead to formation of
shock waves and their subsequent decay. An important point is that for fms waves, this effect works
even when the wave frequency exceeds the plasma frequency, i.e. when the MHD approximation is
generally violated. The dispersion relation for electro-magnetic waves polarized perpendicularly to
the background magnetic field looks like (e.g., \citealt{Melrose97})
 \eqb
 \omega'^2=k'^2c^2-\frac{\omega'^2_p\omega'^2}{\omega'^2_B-\omega'^2}.
 \label{disp}\eqe
This relation is reduced to the dispersion of the fms waves (\ref{fms-disp}) at
$\omega'\ll\omega'_B$ therefore in the magnetically dominated plasma, $\omega_B\gg\omega_p$, this
wave behaves as an fms wave even at $\omega'_p<\omega'<\omega'_B$, when the MHD approximation is
generally inapplicable. Therefore the non-linear steepening proceeds until the the cyclotron
resonance radius (\ref{cycl_radius}).
 The shock formation scale for a strong pulse of a width $l$  is given in the comoving plasma frame
 by eq. (\ref{steepening}). For waves with the amplitude smaller than the background
 field this equation is modified to
 \eqb
  s'_{\rm steep}\sim \sigma \frac{B'}{\delta B'}\frac c{\omega'}.
\eqe
Transforming to the lab frame and making use of eqs. (\ref{energy-total}),  (\ref{energy}) and
(\ref{gamma_pulse}), one finds that the decay of the fms waves due to the non-linear steepening
could occur at the distances
 \eqb
R_{\rm steep}=4\sigma_{\rm pulse}\gamma^2_{\rm pulse}\frac{\cal E}{\epsilon_{\rm FRB}}\frac
c{\omega}= 3\cdot 10^{14}\sigma_{\rm pulse}\frac{L_{\rm
pulse,47}\tau_{-3}P}{f_{-2}\mu_{33}^{10/9}{\cal M}_3^{2/3}\nu_9}\,\rm cm.
 \eqe
The waves survive if this distance exceeds the cyclotron resonance distance (\ref{cycl_radius}),
which implies a not very restrictive condition $\sigma_{\rm pulse}>10$.

 Now let us consider the {\it induced scattering}. Within the cyclotron resonance
radius, there is no scattering at all because the particles experience only drift oscillations in
the crossed electric field of the waves and the magnetic field of the background. Inasmuch as the
drift motion is independent of the particle charge, there is no oscillating electric current that
produces scattered waves. Beyond the cyclotron resonance, $R>R_{\rm cycl}$, the induced scattering
 may be estimated from the Kompaneets equation
 \eqb
\frac{\partial n_{\nu}}{\partial t}=\frac{\sigma_TN'h}{m_ec^2}\frac
1{\nu'^2}\frac{\partial}{\partial\nu'}\left(\nu'^4n^2_{\nu}\right).
 \eqe
 Here $\sigma_T$ is the Thomson cross section, $h$ the Plank constant, $n_{\nu}$ the photon occupation number.
One sees  that the induced scattering rate may be presented, in the comoving plasma frame, as
 \eqb
q'\sim\sigma_TN'c\frac{h\nu'}{m_ec^2} n_{\nu}\sim\frac{\omega'^4_p}{\omega'^3}\frac{U'_{\rm
rad}}{m_ec^2N'},
 \eqe
 where $U'_{\rm rad}\sim h\nu^3n_{\nu}/c^3$ is the radiation energy density. Making use of the
 relations $\omega'^2B=\sigma\omega'^2_p$ and
 \eqb
\frac{U'_{\rm rad}}{m_ec^2N'}=\frac{U'_{\rm rad}\sigma_{\rm pulse}}{B'^2_{\rm pulse}/4\pi}=\frac
12\frac{\varepsilon_{\rm FRB}\sigma_{\rm pulse}}{\cal E},
 \eqe
 transforming the frequencies to the lab frame and making use of eqs. (\ref{Bpulse}), (\ref{energy-total}) and
(\ref{energy}), one gets the estimate for  the induced scattering optical depth
 \eqb
q't'\sim\frac{\omega^4_B}{\omega^3\gamma_{\rm pulse}\sigma_{\rm pulse}^2}\frac{\varepsilon_{\rm
FRB}\sigma_{\rm pulse}}{\cal E}\frac R{c\gamma_{\rm pulse}}=\frac{\omega^4_B}{\sigma_{\rm
pulse}\omega^3}\frac{\varepsilon_{\rm FRB}}{\cal E}\frac R{c\gamma^2_{\rm pulse}}=
\frac{60}{\sigma_{\rm pulse}R^3_{16}}\frac{f_{-2}\mu_{33}^{10/9}{\cal M}_3^{2/3}L_{\rm
pulse}}{\tau_{-3}\nu_9P}\left(\frac{\gamma_{\rm wind}}{\tilde{\gamma_{\rm wind}}}\right)^2.
 \eqe
One sees that the induced scattering may be neglected at a moderately large magnetization,
$\sigma_{\rm pulse}\gtrsim 100$ .

One concludes that at quite general conditions, fms waves could  escape as radio waves .

\section{Polarization of the outgoing radiation}

Let us now consider polarization of the outgoing radiation. In this paper, the source of radiation
is assumed to be the magnetic reconnection producing fms waves. In the fms wave, the electric field
is directed perpendicularly to the
background magnetic field. When 
an fms wave propagates towards decreasing plasma density, it is converted into the electro-magnetic
wave polarized perpendicularly to the local magnetic field.  The waves are generated in the
reconnection process triggered by a large scale magnetic pulse produced by a magnetar flare and
propagating outwards in the magnetar magnetosphere and the wind. The fms waves propagate on the top
of the pulse and escape as radio waves far away from the magnetar, when the pulse is decelerated
colliding with the surrounding plasma. The magnetic pulse is formed well within the magnetosphere
(see eq. \ref{R0}), and the magnetic flux from this region is transferred by the pulse outwards.
Therefore the direction of the magnetic field within the pulse is determined by the rotational
phase of the magnetar. When the pulse travels in the magnetar wind, it picks up the azimuthal
magnetic field, which is accumulated at the front of the pulse (see the discussion at the end of
sect.\ 3).
The radiation passes through this layer on the way out.  If the polarization vector is
adiabatically adjusted to the local magnetic field, the outgoing radiation becomes eventually
completely polarized perpendicularly to the field in the wind, i.e. along the rotational axis of
the magnetar. In the opposite case, the waves are split into two normal modes corresponding to the
local magnetic field so that the radiation is depolarized.


 In the comoving plasma frame, the condition for adiabatic adjustment of the polarization vector
is written as (e.g., \citealt{Ginzburg70})
 \eqb
\frac{\omega'}c\Delta n\delta'>1,
 \label{adiab}\eqe
 where $\delta$ is the characteristic width of the polarization vector rotation zone, $\Delta n$ the difference
 between the refraction indexes of the two normal modes in the system.
In the electron-positron plasma, the modes are linearly polarized. One of them is polarized
perpendicularly to the background magnetic field and is described by the dispersion law
(\ref{disp}). The second mode is polarized along the magnetic field  and therefore it does not
"feel" the magnetic field; it is described by the usual dispersion law,
$\omega'^2=k'^2c^2+\omega'^2_p$. Taking into account that $\omega\lesssim\omega_B$ in the wave
escape zone one could take $\Delta n\sim (\omega'_p/\omega')^2$.

The minimal width of the transition zone where the background magnetic field turns, may be
estimated from the condition that there is enough charge carriers in order to provide the required
curl of the magnetic field,
\eqb
 \frac{B'}{\delta'}< 4\pi eN'.
\eqe
Substituting this inequality into the lhs of the inequality (\ref{adiab}), one gets
 \eqb
\frac{\omega'}c\Delta n\delta'>\frac{\omega'^2_p}{c\omega'}\frac{B'}{4\pi e
N'}=\frac{\omega'_B}{\omega'}.
 \eqe
One sees that if the radiation escapes the magnetic pulse before the cyclotron resonance radius
(\ref{cycl_radius}) is reached, the 100\% polarized emission could be formed. This happens if the
magnetar wind termination shock lies within the cyclotron resonance radius or if the pulse collides
with the mildly relativistic baryonic material ejected by the previous flare. The last is relevant
only for repeating FRBs. An essential point is that in this case, the polarization is directed
along the rotation axis of the magnetar, and therefore the position angle does not change from one
burst to another. If the waves escape the magnetic pulse beyond the cyclotron resonance radius, the
radiation is depolarized, the final degree of polarization being dependent on the angle between the
magnetic field in the pulse and in the wind.

\section{Conclusions}

In this paper, I outlined a model, which attributes the radio emission of FRBs to the magnetic
reconnection during the magnetar flare. The work was motivated by the discovery of 100\%
polarization in a few FRBs \citep{Gijjar18,Michili18,oslowski19},  which imposes severe
restrictions on the emission process because one of the two polarization modes should be suppressed
completely. The reconnection is accompanied by generation of an fms noise, which could be converted
into the electro-magnetic radiation if the magnetic energy in the system exceeds the plasma energy.
The fms waves polarized perpendicularly to the background magnetic field, and they are converted
into a 100\% polarized electro-magnetic waves provided depolarization processes do not come into
play.

One can expect the violent reconnection at the onset of the magnetar flare when a strong, large
scale magnetic pulse from the flare reaches the current sheet separating, just beyond the light
cylinder, the oppositely directed magnetic fields.  It was shown in the paper that the energy and
characteristic frequency of the generated fms waves are compatible with the observed parameters of
FRBs. These waves propagate outwards on the top of the  magnetic pulse. I considered different
processes, which could hinder propagation of the waves on the way out, such as cyclotron
absorption, non-linear wave interactions, induced scattering, and showed that the wave could escape
as electro-magnetic waves. The waves are polarized perpendicularly to the background magnetic field
and, unless the depolarization occurs, the outgoing radiation acquires 100\% polarization degree.

\section*{Acknowledgments}

This research was supported by the grant I-1362-303.7/2016 from the German-Israeli Foundation for
Scientific Research and Development and by the grant 2067/19 from the Israeli Science Foundation.

\appendix
\section{A. FMS waves in the inner magnetosphere}

At the onset of the magnetar flare, the violent reconnection occurs in the inner part of the
magnetosphere therefore one can expect that fms waves are generated there. The plasma scales in the
inner magnetosphere are microscopically small therefore it is not evident that the waves in the GHz
frequency band could be generated. In this appendix, I show that even if we assume that the fms
waves with the required frequencies and power are produced deep inside the magnetosphere, they
would decay via the non-linear interactions (\ref{fms_decay}) so that no radiation would escape.

The non-linear interaction rate is given by eq. (\ref{q}). Denoting the luminosity of the FMS waves
as $L_{\rm FRB}$, one gets the estimate for the wave amplitudes,
\eqb
\delta B=\sqrt{\frac{L_{\rm FRB}}c}\frac 1R,
 \label{amplitude}\eqe
so that
 \eqb
\frac{\delta B}B=\sqrt{\frac{L_{\rm FRB}}c}\frac{R^2}{\mu}.
 \label{relativ-ampl}\eqe
The MHD cascade develops at the condition (\ref{NLcondition}), which in our case is written just as
 \eqb
\left(\frac{\delta B}B\right)^2\omega\frac Rc<10.
 \eqe
  Substituting the above estimate, one finds that fms waves could not propagate beyond the distance
 \eqb
R=1.1\cdot 10^7\left(\frac{\mu_{33}^2}{L_{\rm FRB, 43}\nu_9}\right)^{1/5}\,\rm cm,
 \eqe
which is well within the magnetosphere. Note that the magnetospheric magnetic field exceeds the
perturbation field (\ref{Bpulse}) at this distance, so that the use of the dipole formula for the
background field in the estimate (\ref{relativ-ampl}) is justified.

\section{B. Propagation of a magnetic pulse through the magnetar wind}

 The magnetic perturbation from a magnetar flare propagates outwards as a large scale fms
pulse. The propagation of a non-linear wave in a highly magnetized wind is described as follows
\citep{Lyubarsky03}. In the far zone of the wind, the magnetic field may be considered as purely
azimuthal and the pulse  as spherical. It follows from the continuity equation,
\eqb
\frac{\partial}{\partial t}N+\frac 1{R^2}\frac{\partial}{\partial R} R^2Nv=0, \label{cont}
\eqe
and the frozen-in condition,
 \eqb \frac{\partial B}{\partial t}+\frac
1R\frac{\partial}{\partial R}RvB=0,
\eqe
that the magnetic field may be presented as
\eqb
B=bRN,
 \label{b}\eqe
 where $b$ is a constant. The dynamic equation may be presented in the
form of the energy equation
\eqb
\frac{\partial }{\partial t}T_{00}+\frac 1{R^2}\frac{\partial }{\partial R}R^2T_{01}=0,
\eqe
where the components of the energy-momentum tensor are
\begin{eqnarray}
T_{01}=w\gamma^2v+\frac{B^2}{4\pi}\gamma^2 v, \\
 T_{00}=w\gamma^2
-p+\frac{1+v^2}{8\pi}B^2\gamma^2.\label{T00}
\end{eqnarray}
Here $w$ and $p$ are the plasma enthalpy and the pressure, correspondingly.

When the  plasma is cold, $w=mN'c^2$, $p=0$, where the prime is referred to the plasma comoving
frame, one can define the new variable, \eqb
\tilde{N}=NR^2,
 \label{tildaN}\eqe
 and write eqs. (\ref{cont}-\ref{T00}) as a
set of the 1D hydrodynamics equations
\begin{eqnarray}
\frac{\partial\tilde{N}}{\partial t}+\frac{\partial v\tilde{N}}{\partial R}=0,\label{cont_eq}\\
\frac{\partial T_{00}}{\partial t}+\frac{\partial T_{01}}{\partial R}=0,\label{energy_eq}\\
T_{01}={\cal W}\gamma^2v;\qquad
 T_{00}={\cal W}\gamma^2-\cal{P},\\
 \end{eqnarray}
for the medium with the effective equations of state and energy
 \begin{eqnarray}
 {\cal P}=\frac{R^2B'^2}{8\pi}=\frac{b^2\tilde{N'}^2}{8\pi}; \qquad {\cal E}=m\tilde{N'}c^2+\frac{b^2\tilde{N'}^2}{8\pi};\\
{\cal W}={\cal E}+{\cal P}=m\tilde{N'}c^2+\frac{b^2\tilde{N'}^2}{4\pi}.
\end{eqnarray}
Now the solution may be found in the form of a simple wave (e.g., \citealt{Landau_Lifshitz87}).

The nonlinear simple wave is found from the condition that all dependent variables are functions of
one of them, e.g., $\tilde{N}$. This means that equations (\ref{cont_eq}) and (\ref{energy_eq})
should be equivalent, which implies,
\eqb
\frac{dT_{01}}{dT_{00}}=\frac{d(\tilde{N}v)}{d(\tilde{N})}.
 \eqe
  The last equation reduces to
\eqb
\frac{dv}{d\tilde{N}}=\frac{s_{\rm fms}}{\gamma^2\tilde{N}(1+vs_{\rm fms}/c^2)},
 \label{dvdN}\eqe
 where
 \eqb
s_{\rm fms}=\left(\frac{d\cal P}{d\cal E}\right)^{1/2}c=\left(\frac{b^2\tilde{N}^2}{4\pi
m\tilde{N}\gamma c^2+b^2\tilde{N}^2}\right)^{1/2}c
 \eqe
is the fms velocity. Making use of eqs. (\ref{b}) and (\ref{tildaN}), one presents $s_{\rm fms}$ in
the standard form:
 \eqb
s_{\rm fms}=\left(\frac{B^2}{4\pi mN\gamma
c^2+B^2}\right)^{1/2}c=\left(\frac{\sigma}{1+\sigma}\right)^{1/2}c,
 \eqe
where $\sigma$ is the magnetization parameter. For a small wave amplitude, $\delta \tilde{n}\ll
\tilde{n}$, one obtains the linear FMS wave propagating with the velocity $s_{\rm fms}=\it const$.
For a high amplitude pulse, $s_M$ depends on the local density therefore the wave becomes
nonlinear. However in the strongly magnetized case, $\sigma\gg 1$, $s_{\rm fms}$ is close to the
speed of light, therefore even  high amplitude waves are nearly linear. The physical reason for
this is that the displacement current significantly exceeds the conductivity current (see eq.
\ref{displacement}).

Substituting $s_{\rm fms}=c$ and  $v=\left(1-1/(2\gamma^2)\right)c$ into eq.\ (\ref{dvdN}), one
gets
 \eqb
\frac{d\gamma}{d\tilde{N}}=\frac{\gamma}{2\tilde{N}},
 \eqe
 which yields
 \eqb
\gamma=\left(\frac{\tilde{N}}{\tilde{N}_{\rm wind}}\right)^{1/2}\gamma_{\rm wind}=
\left(\frac{B}{B_{\rm wind}}\right)^{1/2}\gamma_{\rm wind}.
 \eqe
Here the index "wind" is referred to quantities in the magnetar wind outside of the pulse.
Note that $v$ and $\gamma$ are the velocity and the Lorentz factor of the plasma in the pulse,
i.e., of the zero electric field frame (see eqs.\ \ref{Gamma} end \ref{gamma_pulse})). The pulse
itself, i.e. the waveform, moves along the characteristics of eqs.\ (\ref{cont_eq},
\ref{energy_eq})
 \eqb
\frac{dx}{dt}=\frac{d(N v)}{dN}=\frac{v+s_{\rm fms}}{1+vs_{\rm fms}/c^2}= \left(1-\frac
1{8\gamma^2\sigma}\right)c.
 \eqe
 The last equality in this expression is obtained for $\sigma\gg 1$, $\gamma\gg1$. One sees
 that the waveform moves with the Lorentz factor $2\gamma\sqrt{\sigma}$, so that at each point, the waveform moves with
respect to the plasma with the local fms velocity. Inasmuch as the velocity
 of  the waveform varies along the pulse, the pulse shape also varies,
 which could lead to formation of a shock front. However, the corresponding distance, $\sim
 8\gamma^2\sigma l$, is so large that for the parameters of the pulse discussed in sect.\ 3,
 the non-linear steepening could be neglected for any reasonable distances from the
 magnetar. Therefore one can safely assume that the pulse moves with the speed of light.


\begin{thebibliography}{43}
\expandafter\ifx\csname natexlab\endcsname\relax\def\natexlab#1{#1}\fi

\bibitem[{{Appl} \& {Camenzind}(1988)}]{Appl_Camenzind88}
{Appl}, S., \& {Camenzind}, M. 1988, \aap, 206, 258

\bibitem[{{Beloborodov}(2013)}]{Beloborodov13a}
{Beloborodov}, A.~M. 2013, \apj, 762, 13

\bibitem[{{Beloborodov}(2017)}]{Beloborodov17}
---. 2017, \apjl, 843, L26

\bibitem[{{Beloborodov}(2019)}]{Beloborodov19}
---. 2019, arXiv e-prints, arXiv:1908.07743

\bibitem[{{Beloborodov} \& {Thompson}(2007)}]{BeloborodovThompson07}
{Beloborodov}, A.~M., \& {Thompson}, C. 2007, \apj, 657, 967

\bibitem[{{Beskin} {et~al.}(1998){Beskin}, {Kuznetsova}, \&
  {Rafikov}}]{Beskin98}
{Beskin}, V.~S., {Kuznetsova}, I.~V., \& {Rafikov}, R.~R. 1998, \mnras, 299,
  341

\bibitem[{{Carrasco} {et~al.}(2019){Carrasco}, {Vigan{\`o}}, {Palenzuela}, \&
  {Pons}}]{Carrasco_etal19}
{Carrasco}, F., {Vigan{\`o}}, D., {Palenzuela}, C., \& {Pons}, J.~A. 2019,
  \mnras, 484, L124

\bibitem[{{Cordes} \& {Wasserman}(2016)}]{Cordes_Wasserman16}
{Cordes}, J.~M., \& {Wasserman}, I. 2016, \mnras, 457, 232

\bibitem[{{Gajjar} {et~al.}(2018){Gajjar}, {Siemion}, {Price}, {Law},
  {Michilli}, {Hessels}, {Chatterjee}, {Archibald}, {Bower}, {Brinkman},
  {Burke-Spolaor}, {Cordes}, {Croft}, {Enriquez}, {Foster}, {Gizani},
  {Hellbourg}, {Isaacson}, {Kaspi}, {Lazio}, {Lebofsky}, {Lynch}, {MacMahon},
  {McLaughlin}, {Ransom}, {Scholz}, {Seymour}, {Spitler}, {Tendulkar},
  {Werthimer}, \& {Zhang}}]{Gijjar18}
{Gajjar}, V., {Siemion}, A.~P.~V., {Price}, D.~C., {Law}, C.~J., {Michilli},
  D., {Hessels}, J.~W.~T., {Chatterjee}, S., {Archibald}, A.~M., {Bower},
  G.~C., {Brinkman}, C., {Burke-Spolaor}, S., {Cordes}, J.~M., {Croft}, S.,
  {Enriquez}, J.~E., {Foster}, G., {Gizani}, N., {Hellbourg}, G., {Isaacson},
  H., {Kaspi}, V.~M., {Lazio}, T.~J.~W., {Lebofsky}, M., {Lynch}, R.~S.,
  {MacMahon}, D., {McLaughlin}, M.~A., {Ransom}, S.~M., {Scholz}, P.,
  {Seymour}, A., {Spitler}, L.~G., {Tendulkar}, S.~P., {Werthimer}, D., \&
  {Zhang}, Y.~G. 2018, \apj, 863, 2

\bibitem[{{Gallant} {et~al.}(1992){Gallant}, {Hoshino}, {Langdon}, {Arons}, \&
  {Max}}]{Gallant92}
{Gallant}, Y.~A., {Hoshino}, M., {Langdon}, A.~B., {Arons}, J., \& {Max}, C.~E.
  1992, \apj, 391, 73

\bibitem[{{Ghisellini}(2017)}]{Gisellini17}
{Ghisellini}, G. 2017, \mnras, 465, L30

\bibitem[{{Ghisellini} \& {Locatelli}(2018)}]{Gisellini_Locatelli18}
{Ghisellini}, G., \& {Locatelli}, N. 2018, \aap, 613, A61

\bibitem[{{Gil} {et~al.}(2004){Gil}, {Lyubarsky}, \& {Melikidze}}]{Gil_etal04}
{Gil}, J., {Lyubarsky}, Y., \& {Melikidze}, G.~I. 2004, \apj, 600, 872

\bibitem[{{Gill} {et~al.}(2018){Gill}, {Granot}, \& {Lyubarsky}}]{Gill_etal18}
{Gill}, R., {Granot}, J., \& {Lyubarsky}, Y. 2018, \mnras, 474, 3535

\bibitem[{{Ginzburg}(1970)}]{Ginzburg70}
{Ginzburg}, V.~L. 1970, {The propagation of electromagnetic waves in plasmas}
  (Pergamon Press, Oxford)

\bibitem[{{Gruzinov} \& {Waxman}(2019)}]{Gruzinov_Waxman19}
{Gruzinov}, A., \& {Waxman}, E. 2019, \apj, 875, 126

\bibitem[{{Kagan} {et~al.}(2015){Kagan}, {Sironi}, {Cerutti}, \&
  {Giannios}}]{Kagan_etal15}
{Kagan}, D., {Sironi}, L., {Cerutti}, B., \& {Giannios}, D. 2015, \ssr, 191,
  545

\bibitem[{{Kaspi} \& {Beloborodov}(2017)}]{Kaspi_Beloborodov17}
{Kaspi}, V.~M., \& {Beloborodov}, A.~M. 2017, \araa, 55, 261

\bibitem[{{Katz}(2018)}]{Katz18}
{Katz}, J.~I. 2018, \mnras, 481, 2946

\bibitem[{{Kirk} \& {Skj{\ae}raasen}(2003)}]{Kirk_Skjeraasen03}
{Kirk}, J.~G., \& {Skj{\ae}raasen}, O. 2003, \apj, 591, 366

\bibitem[{{Kumar} {et~al.}(2017){Kumar}, {Lu}, \& {Bhattacharya}}]{Kumar17}
{Kumar}, P., {Lu}, W., \& {Bhattacharya}, M. 2017, \mnras, 468, 2726

\bibitem[{{Landau} \& {Lifshitz}(1987)}]{Landau_Lifshitz87}
{Landau}, L.~D., \& {Lifshitz}, E.~M. 1987, {Fluid Mechanics}

\bibitem[{{Lu} \& {Kumar}(2019)}]{Lu_Kumar19}
{Lu}, W., \& {Kumar}, P. 2019, \mnras, 483, L93

\bibitem[{{Lyubarskii}(1996)}]{Lyubarskii96}
{Lyubarskii}, Y.~E. 1996, \aap, 311, 172

\bibitem[{{Lyubarsky}(2010)}]{Lyubarsky10}
{Lyubarsky}, Y. 2010, \apjl, 725, L234

\bibitem[{{Lyubarsky}(2014)}]{Lyubarsky14}
---. 2014, \mnras, 442, L9

\bibitem[{{Lyubarsky}(2019)}]{Lyubarsky19}
---. 2019, \mnras, 483, 1731

\bibitem[{{Lyubarsky} \& {Kirk}(2001)}]{Lyubarsky_Kirk01}
{Lyubarsky}, Y., \& {Kirk}, J.~G. 2001, \apj, 547, 437

\bibitem[{{Lyubarsky}(2003)}]{Lyubarsky03}
{Lyubarsky}, Y.~E. 2003, \mnras, 339, 765

\bibitem[{{Margalit} {et~al.}(2019){Margalit}, {Metzger}, \&
  {Sironi}}]{Metzger_etal20}
{Margalit}, B., {Metzger}, B.~D., \& {Sironi}, L. 2019, arXiv e-prints,
  arXiv:1911.05765

\bibitem[{{Melrose}(1997)}]{Melrose97}
{Melrose}, D.~B. 1997, Plasma Physics and Controlled Fusion, 39, A93

\bibitem[{{Metzger} {et~al.}(2019){Metzger}, {Margalit}, \&
  {Sironi}}]{Metzger19}
{Metzger}, B.~D., {Margalit}, B., \& {Sironi}, L. 2019, \mnras, 485, 4091

\bibitem[{{Michilli} {et~al.}(2018){Michilli}, {Seymour}, {Hessels}, {Spitler},
  {Gajjar}, {Archibald}, {Bower}, {Chatterjee}, {Cordes}, {Gourdji}, {Heald},
  {Kaspi}, {Law}, {Sobey}, {Adams}, {Bassa}, {Bogdanov}, {Brinkman},
  {Demorest}, {Fernand ez}, {Hellbourg}, {Lazio}, {Lynch}, {Maddox}, {Marcote},
  {McLaughlin}, {Paragi}, {Ransom}, {Scholz}, {Siemion}, {Tendulkar}, {van
  Rooy}, {Wharton}, \& {Whitlow}}]{Michili18}
{Michilli}, D., {Seymour}, A., {Hessels}, J.~W.~T., {Spitler}, L.~G., {Gajjar},
  V., {Archibald}, A.~M., {Bower}, G.~C., {Chatterjee}, S., {Cordes}, J.~M.,
  {Gourdji}, K., {Heald}, G.~H., {Kaspi}, V.~M., {Law}, C.~J., {Sobey}, C.,
  {Adams}, E.~A.~K., {Bassa}, C.~G., {Bogdanov}, S., {Brinkman}, C.,
  {Demorest}, P., {Fernand ez}, F., {Hellbourg}, G., {Lazio}, T.~J.~W.,
  {Lynch}, R.~S., {Maddox}, N., {Marcote}, B., {McLaughlin}, M.~A., {Paragi},
  Z., {Ransom}, S.~M., {Scholz}, P., {Siemion}, A.~P.~V., {Tendulkar}, S.~P.,
  {van Rooy}, P., {Wharton}, R.~S., \& {Whitlow}, D. 2018, \nat, 553, 182

\bibitem[{{Os{\l}owski} {et~al.}(2019){Os{\l}owski}, {Shannon}, {Ravi},
  {Kaczmarek}, {Zhang}, {Hobbs}, {Bailes}, {Russell}, {van Straten}, {James},
  {Jameson}, {Mahony}, {Kumar}, {Andreoni}, {Bhat}, {Burke-Spolaor}, {Dai},
  {Dempsey}, {Kerr}, {Manchester}, {Parthasarathy}, {Reardon}, {Sarkissian},
  {Spiewak}, {Toomey}, {Wang}, {Zhang}, \& {Zhu}}]{oslowski19}
{Os{\l}owski}, S., {Shannon}, R.~M., {Ravi}, V., {Kaczmarek}, J.~F., {Zhang},
  S., {Hobbs}, G., {Bailes}, M., {Russell}, C.~J., {van Straten}, W., {James},
  C.~W., {Jameson}, A., {Mahony}, E.~K., {Kumar}, P., {Andreoni}, I., {Bhat},
  N.~D.~R., {Burke-Spolaor}, S., {Dai}, S., {Dempsey}, J., {Kerr}, M.,
  {Manchester}, R.~N., {Parthasarathy}, A., {Reardon}, D., {Sarkissian}, J.~M.,
  {Spiewak}, R., {Toomey}, L., {Wang}, J.~B., {Zhang}, L., \& {Zhu}, X.~J.
  2019, \mnras, 488, 868

\bibitem[{{Parfrey} {et~al.}(2013){Parfrey}, {Beloborodov}, \&
  {Hui}}]{Parfrey_etal13}
{Parfrey}, K., {Beloborodov}, A.~M., \& {Hui}, L. 2013, \apj, 774, 92

\bibitem[{{Philippov} {et~al.}(2019){Philippov}, {Uzdensky}, {Spitkovsky}, \&
  {Cerutti}}]{Philippov19}
{Philippov}, A., {Uzdensky}, D.~A., {Spitkovsky}, A., \& {Cerutti}, B. 2019,
  \apjl, 876, L6

\bibitem[{{Thompson}(2008)}]{Thompson08}
{Thompson}, C. 2008, \apj, 688, 1258

\bibitem[{{Thompson} \& {Blaes}(1998)}]{Thompson_Blaes98}
{Thompson}, C., \& {Blaes}, O. 1998, \prd, 57, 3219

\bibitem[{{Thompson} {et~al.}(2002){Thompson}, {Lyutikov}, \&
  {Kulkarni}}]{ThompsonLyutikovKulkarni02}
{Thompson}, C., {Lyutikov}, M., \& {Kulkarni}, S.~R. 2002, \apj, 574, 332

\bibitem[{{Uzdensky} \& {Spitkovsky}(2014)}]{Uzdensky_Spitkovsky14}
{Uzdensky}, D.~A., \& {Spitkovsky}, A. 2014, \apj, 780, 3

\bibitem[{{Wang} {et~al.}(2019){Wang}, {Zhang}, {Chen}, \& {Xu}}]{Wang_etal19}
{Wang}, W., {Zhang}, B., {Chen}, X., \& {Xu}, R. 2019, \apj, 875, 84

\bibitem[{{Waxman}(2017)}]{Waxman17}
{Waxman}, E. 2017, \apj, 842, 34

\bibitem[{{Yang} \& {Zhang}(2018)}]{Yang_Zhang18}
{Yang}, Y.-P., \& {Zhang}, B. 2018, \apj, 868, 31

\end{thebibliography}


\end{document}